\documentclass[a4paper]{article}
\usepackage{amscd,amsmath,amssymb}
\parskip=\medskipamount

\newcommand{\p}[1]{(\ref{#1})}

\newcommand{\bQ}{{\overline Q}{}}

\newcommand{\bpsi}{{\bar\psi}{}}
\newcommand{\brho}{{\bar\rho}{}}

\newcommand{\bv}{{\bar v}}

\newcommand{\cN}{{\cal N}}
\newcommand{\Nf}{{{\cal N}{=}\,4}}

\newcommand{\und}{\qquad\textrm{and}\qquad}

\newcommand{\be}{\begin{equation}}
\newcommand{\ee}{\end{equation}}
\newcommand{\bea}{\begin{eqnarray}}
\newcommand{\eea}{\end{eqnarray}}

\newcommand{\ba}{\begin{array}} \newcommand{\ea}{\end{array}}

\def\im{{\rm i}}
\def\sfrac#1#2{{\textstyle\frac{#1}{#2}}}

\begin{document}
\thispagestyle{empty}
\begin{flushright}
\end{flushright}\vspace{1cm}
\begin{center}
{\LARGE\bf $\cN$-extended supersymmetric Calogero models }
\end{center}
\vspace{1cm}

\begin{center}
{\Large\bf  Sergey Krivonos${}^a$, Olaf Lechtenfeld$^b$ and Anton Sutulin${}^a$}
\end{center}

\vspace{0.2cm}

\begin{center}
{${}^a$ \it
Bogoliubov  Laboratory of Theoretical Physics, JINR,
141980 Dubna, Russia}

${}^b$ {\it
Institut f\"ur Theoretische Physik and Riemann Center for Geometry and Physics \\
Leibniz Universit\"at Hannover,
Appelstrasse 2, D-30167 Hannover, Germany}

\vspace{0.5cm}

{\tt krivonos@theor.jinr.ru, lechtenf@itp.uni-hannover.de, sutulin@theor.jinr.ru}
\end{center}
\vspace{2cm}

\begin{abstract}
\noindent
We propose a new $\cN$-extended supersymmetric $su(n)$ spin-Calogero model. 
Employing a generalized Hamiltonian reduction adopted to the supersymmetric case, 
we explicitly construct a novel rational $n$-particle Calogero model with an arbitrary even number of supersymmetries.
It features $\cN n^2$ rather than $\cN n$ fermionic coordinates and increasingly high fermionic powers
in the supercharges and the Hamiltonian.
\end{abstract}

\newpage
\setcounter{page}{1}
\setcounter{equation}{0}
\section{Introduction}
The original rational Calogero model of $n$ interacting identical particles on a line~\cite{Calogero}, 
pertaining to the roots of $A_1\oplus A_{n-1}$ and given by the classical Hamiltonian
\be\label{CH}
H=\frac{1}{2} \sum_{i=1}^n p_i^2 + \frac{1}{2}\sum_{i\neq j} \frac{g^2}{\left( x^i{-}x^j\right)^2}\ ,
\ee
has often been the subject of ``supersymmetrization''. In this endeavor, {\it extended\/} supersymmetry has turned out to be surprisingly rich.
After the straightforward formulation of $\cN{=}\,2$ supersymmetric Calogero models by Freedman and Mende~\cite{FM},
a barrier was encountered at $\Nf$~\cite{W}.
An important step forward then was the explicit construction of the supercharges and the Hamiltonian for the $\Nf$ supersymmetric 
three-particle Calogero model~\cite{BGL,GOP}, which introduced a second prepotential $F$ besides the familiar prepotential~$U$.
However, it was found that quantum corrections modify the potential in (\ref{CH}), and that $F$ is subject to intricate nonlinear differential equations, 
the WDVV equations, beyond the three-particle case. These results were then confirmed and elucidated in a superspace description~\cite{BKS}.
Finally, extending the system by a single harmonic degree of freedom ($su(2)$ spin variables~\cite{FIL}) it was possible to
write down a unique $osp(4|2)$ symmetric four-particle Calogero model~\cite{KL}.~\footnote{
Here and in the above history, the goal is a bosonic potential exactly as in (\ref{CH}). 
Models with more general interactions can be found for any number of particles.}
A detailed discussion concerning the supersymmetrization of the Calogero models can be found in the review~\cite{Iv3}.

It seems that a guiding principle was missing for the construction of extended supersymmetric Calogero models.
Indeed, while for $n\leq3$ translation and (super-)conformal symmetry almost completely defines the system, 
the $n\geq4$ cases admit a lot of freedom which cannot {\it a priori\/} be fixed.
In the bosonic case, such a guiding principle exists~\cite{KKS}. The Calogero model as well as its different extensions (see, e.g. \cite{w,GS,gf}) 
are closely related with matrix models and can be obtained from them by a reduction procedure (see \cite{Poly1} for first results and \cite{Poly2} for a review). 
If we want to employ this principle also for finding extended supersymmetric Calogero models, then the two main steps are 
\\[-16pt]
\begin{itemize}
\addtolength{\itemsep}{-6pt}
\item supersymmetrization of a matrix model
\item supersymmetrization of the reduction procedure or proper gauge fixing.
\end{itemize} 
This idea is not new. It has successfully been employed in~\cite{Iv1, Iv2, Iv4, Iv5}.
The resulting supersymmetric systems feature 
\\[-16pt]
\begin{itemize}
\addtolength{\itemsep}{-6pt}
\item a large number of fermions -- far more than the $4n$ fermions expected in an $\Nf$ $n$-particle system within the standard (but unsuccessful!) approach
\item a rather complicated structure of the supercharges and the Hamiltonian, with fermionic polynomials of maximal degree
\item a variety of bosonic potentials, including $su(2)$ {\it spin\/}-Calogero interactions
\end{itemize}
but they do not contain a genuine $\Nf$ supersymmetric Calogero model, i.e.~one with a mere pairwise inverse-square no-spin bosonic potential.

Here we use the same guiding principle and start with the bosonic $su(n)$ spin-Calogero model in the Hamiltonian approach. 
We then provide an $\cN$-extended supersymmetrization of this system. 
It is important that we do {\it not a priori\/} fix a realization for the $su(n)$ generators. 
Finally we generalize the reduction procedure to the $\cN$-extended system and find the first $\cN$-extended supersymmetric Calogero model,
for {\it any\/} even number of supersymmetries.

\newpage

\setcounter{equation}{0}
\section{$\cN$-extended supersymmetric Calogero model}
\subsection{Bosonic Calogero model from hermitian matrices}
It is well known that the rational $n$-particle Calogero model  \cite{Calogero} can be obtained by Hamiltonian reduction
from the hermitian matrix model \cite{KKS, Poly1}. Adapted to our purposes, the procedure reads as follows.
One starts from the $su(n)$ spin generalization~\cite{GS} of the standard Calogero model, as given by
\be\label{Hb1}
H= \frac{1}{2} \sum_{i=1}^n  p_i^2 + \frac{1}{2}\sum_{i \neq j}^n \frac{\ell_{ij} \ell_{ji}}{\left(x^i{-}x^j\right)^2}\ .
\ee
The particles are described by their coordinates $x^i$ and momenta $p_i$ together with their internal degrees of freedom encoded
in the angular momenta $\left( \ell_{ij}\right)^\dagger =\ell_{ji}$ with $\sum_i \ell_{ii}=0$. The non-vanishing Poisson brackets are
\be\label{PB1}
\left\{ x^i, p_j\right\} = \delta^i_j \und 
\left\{ \ell_{ij}, \ell_{km}\right\}  =\im \left( \delta_{im} \ell_{kj}-\delta_{kj}\ell_{im}\right)\ .
\ee
The Hamiltonian \p{Hb1} follows directly from the free hermitian matrix model (for details see~\cite{Poly2}).

To get the standard Calogero Hamiltonian~(\ref{CH}) from \p{Hb1} one has to reduce the angular sector of the latter, in two steps.
Firstly, one (weakly) imposes the constraints
\be\label{con1}
\ell_{11}\approx\ell_{22}\approx\ldots\approx\ell_{nn}\approx 0\ .
\ee
They commute with the Hamiltonian \p{Hb1} and with each other, hence are of first class.
To resolve them one introduces auxiliary complex variables $v_i$ and $\bv_i=\left(v_i\right)^\dagger$ obeying the Poisson brackets
\be\label{PB2}
\left\{ v_i, \bv_j\right\}= -\im\,\delta_{ij}
\ee
and realizes the $su(n)$ generators $\ell_{ij}$ as
\be\label{ell}
{\hat\ell}_{ij} = - v_i \bv_j +\frac{1}{n} \delta_{ij} \sum_k^n v_k \bv_k\ .
\ee
Secondly, passing to polar variables $r_i$ and $\phi_i$ defined as
\be\label{vvb}
v_i = r_i e^{\im \phi_i} \und  \bv_i = r_i e^{- \im \phi_i}
\qquad \Rightarrow \qquad \left\{r_i, \phi_j\right\} = \frac{1}{2 r_i} \delta_{ij}\ ,
\ee
the constraints \p{con1} are resolved by putting
\be\label{sol1}
r_1\approx r_2\approx\ldots\approx r_n\ .
\ee
Plugging this solution into the Hamiltonian \p{Hb1} one may additionally fix $n{-}1$ angles $\phi_i$, say 
\be\label{sol1a}
\phi_1\approx\phi_2\approx\ldots\approx\phi_{n-1}\approx0\ .
\ee
At this stage the $2n$ variables $\{r_i, \phi_i\}$ are reduced to the two variables $r_n$ and $\phi_n$. 
However, the reduced Hamiltonian  does not depend on $\phi_n$ and has the form
\be\label{Hbred}
H_{\mathrm{red}} = \frac{1}{2}  \sum_{i=1}^n  p_i^2 + \frac{1}{2} \sum_{i \neq j}^n \frac{r_n^4}{\left(x^i{-}x^j\right)^2}\ .
\ee
Therefore
\be\label{redB}
\left\{ H_{\mathrm{red}}, r_n \right\}\approx 0 \und r_n^2 \approx \mathrm{const} =: g\ ,
\ee
and the reduced Hamiltonian $H_{\mathrm{red}}$ coincides with the standard $n$-particle rational Calogero Hamiltonian.
We note that in the bosonic case most reduction steps are not needed, because the Hamiltonian \p{Hb1} 
does not depend on the angles $\phi_i$ at all. However, in the supersymmetric case all reduction steps will be important.

In what follows we will construct an $\cN$-extended supersymmetric generalization of the Hamiltonian \p{Hb1} 
and perform the supersymmetric version of the reduction just discussed, finishing with an $\cN$-extended supersymmetric Calogero model,
for $\cN=2M$ and $M=1,2,3,\ldots$.

\subsection{$\cN$-extended supersymmetric $su(n)$ spin-Calogero model}
On the outset we have to clarify what is the minimal number of fermionic variables necessary to realize an $\cN=2M$ supersymmetric extension
of the $su(n)$ spin-Calogero model \p{Hb1}. Clearly, as partners to the bosonic coordinates $x^i$ one needs $\cN n$  fermions 
$\psi_i^a$ and $\bpsi_{i\;a}$ with $a=1,2,\ldots M$.
However, this is not enough to construct $\cN$ supercharges $Q^a$ and $\bQ_b$ which must generate the $\cN{=}\,2M$ superalgebra
\be\label{N4SP}
\left\{ Q^a , \bQ_b \right\} = - 2 \im\, \delta^a_b\, H \und \left\{ Q^a, Q^b \right\}=\left\{ \bQ_a, \bQ_b \right\}=0 \ .
\ee
The reason is simple: to generate the potential term $\sum_{i \neq j}^n\frac{\ell_{ij}\ell_{ji}}{\left( x^i-x^j\right)^2}$ in the Hamiltonian, 
the supercharges $Q^a$ and $\bQ_b$  must contain the terms
\be
\im \sum_{i\neq j}^n\frac{ \ell_{ij}\rho^a_{ji}}{x^i{-}x^j} \und  -\im \sum_{i\neq j}^n \frac{\ell_{ji}\brho_{ij\, a} }{x^i{-}x^j}\ ,
\ee
respectively, where $\rho^a_{ij}$ and $\brho_{ij\,a}$ are some additional fermionic variables. 
These fermions cannot be constructed from $\psi^a_i$ or $\bpsi_{i\, a}$. Hence, we are forced to introduce $\cN n(n{-}1)$ further independent
fermions $\rho^a_{ij}$ and $\brho_{ij\, a}$ subject to $\rho^a_{ii} = \brho_{ii\,a}=0$ for each value of the index $i$. 
In total, we thus utilize $\cN n^2$ fermions of type $\psi$ or $\rho$, which we demand to obey the following Poisson brackets,
\be\label{PBS1}
\left\{ \psi^a_i, \bpsi_{j\,b}\right\} = -\im\,\delta^a_b \delta_{ij}\ , \quad
\left\{ \rho^a_{ij}, \brho_{km\, b}\right\}= -\im\, \delta^a_b  \delta_{im}\delta_{jk}\ ,\quad\textrm{with}\quad 
\left(\rho^a_{ij}\right)^\dagger=\brho_{ji \, a} \quad\textrm{and}\quad \rho^a_{ii}=\brho_{ii \,a}= 0\ .
\ee

The next important ingredient of our construction is the composite object 
\be\label{Pi1}
\Pi_{ij}=\sum_{a=1}^M \Bigl[ \left( \psi^a_i{-}\psi^a_j\right) \brho_{ij\,a}+\left( \bpsi_{i\,a}{-}\bpsi_{j\,a}\right) \rho^a_{ij}+ 
\sum_{k=1}^n \left( \rho^a_{ik}\brho_{kj\,a}+\brho_{ik\,a}\rho^a_{kj}\right)\Bigr]
\qquad\Rightarrow\qquad \left( \Pi_{ij}\right)^\dagger = \Pi_{ji}\ .
\ee
One may check that, with respect to the brackets \p{PBS1}, the $\Pi_{ij}$ form an $su(n)$ algebra just like the $\ell_{ij}$,
\be\label{sunPi}
 \left\{ \Pi_{ij}, \Pi_{km} \right\}=\im \left( \delta_{im} \Pi_{kj}-\delta_{kj}\Pi_{im}\right)\ ,
\ee
and they commute with the our fermions as follows,
\be
\begin{aligned}
& \left\{ \Pi_{ij}, \psi^a_k\right\} = \im \left( \delta_{ik}{-}\delta_{jk}\right) \rho^a_{ij}\ , \qquad
& \left\{ \Pi_{ij}, \rho^a_{km} \right\} = -\im\, \delta_{im}\delta_{jk} \left( \psi^a_i {-}\psi^a_j\right) -
\im \delta_{jk} \rho^a_{im} +\im \delta_{im}\rho^a_{kj}\ , \\[4pt]
& \left\{ \Pi_{ij}, \bpsi_{k\, a}\right\} = \im \left( \delta_{ik}{-}\delta_{jk}\right) \brho_{ij\,a}\ , \qquad
& \left\{ \Pi_{ij}, \brho_{km\, a} \right\} = -\im\, \delta_{im}\delta_{jk} \left( \bpsi_{i\,a} {-}\bpsi_{j\,a}\right) -
\im \delta_{jk} \brho_{im\,a} +\im \delta_{im}\brho_{kj\,a}\ .
\end{aligned}
\ee
It is a matter of straightforward calculation to check that the supercharges 
\be\label{scharges}
Q^a= \sum_{i=1}^n p_i \psi^a_i +\im \sum_{i \neq j}^n \frac{\left( \ell_{ij}+\Pi_{ij}\right) \rho^a_{ji}}{x^i-x^j} \und
\bQ_b= \sum_{i=1}^n p_i \bpsi_{i\,b} - \im \sum_{i \neq j}^n \frac{\brho_{ij\,b}\left( \ell_{ji}+\Pi_{ji}\right) }{x^i-x^j}
\ee
obey the $\cN{=}\,2M$ superalgebra \p{N4SP} with the Hamiltonian
\be\label{N4Ham}
H= \frac{1}{2}\sum_{i=1}^n p_i^2 + \frac{1}{2}\sum_{i \neq j}^n 
\frac{\left( \ell_{ij}+\Pi_{ij}\right)\left( \ell_{ji}+\Pi_{ji}\right) }{\left(x^i-x^j\right)^2}\ ,
\ee
modulo the first-class constraints
\be\label{scon1}
\chi_i:=\ell_{ii}+\Pi_{ii} \approx  0 \quad \forall\; i\ , 
\ee
with
\be
\left\{Q^a, \chi_i\right\}\approx\left\{\bQ_a, \chi_i\right\}\approx\left\{H, \chi_i\right\} \approx\left\{\chi_i, \chi_j\right\}\approx 0\ .
\ee
The supercharges $Q^a$ and $\bQ_b$ in~\p{scharges} and the Hamiltonian $H$ in~\p{N4Ham} describe the $\cN{=}\,2M$ supersymmetric $su(n)$ spin-Calogero model.

For $\Nf$ it essentially coincides with the $osp(4|2)$ supersymmetric mechanics  constructed in \cite{Iv1,Iv2}.
However, there are a few differences:
\\[-16pt]
\begin{itemize}
\addtolength{\itemsep}{-6pt}
\item The Hamiltonian \p{N4Ham} has no interaction for the center-of-mass coordinate $X=\sum_i x^i$.
Correspondingly, the supercharges \p{scharges} do not include certain terms which appeared in \cite{Iv1,Iv2}.
\item Working at the Hamiltonian level, we may keep the $su(n)$ generators~$\ell_{ij}$ unspecified. Precisely this
enables the minimal realization \p{ell} with a minimal number of auxiliary variables $v_i, \bv_i$.
At the Lagrangian level this corresponds to using $(2,4,2)$ supermultiplets for the auxiliary bosonic superfields 
instead of $(4,4,0)$ superfields as in~\cite{Iv1,Iv2}.
\end{itemize}
Now we are ready to reduce our $\cN{=}\,2M$ $su(n)$ spin-Calogero model to a genuine $\cN{=}\,2M$ Calogero model.

\subsection{$\cN$-extended supersymmetric (no-spin) Calogero models}
As we can see from the previous subsection, the supersymmetric analogs \p{scon1} of the purely bosonic constraints \p{con1} appear automatically.  
These constraints generate $n{-}1$ local U(1) transformations\footnote{
Due to the relation $\sum_i^n \chi_i =0$ we have only $n{-}1$ independent constraints.} 
of the variables $\{v_i, \bv_i, \rho^a_{ij}, \brho_{ij\,a}\}$. 
In terms of the $2n$ polar variables $r_i$ and $\phi_i$ defined in~\p{vvb}, the constraints \p{scon1} can be easily resolved as
\be\label{solcon}
r^2_k \approx r^2_n+\Pi_{kk}-\Pi_{nn} \qquad\mathrm{for}\quad k=1,\ldots, n{-}1\ .
\ee
After fixing the residual gauge freedom as
\be\label{gf1}
\phi_1\approx\phi_2\approx\ldots\approx\phi_{n-1} \approx 0\ ,
\ee
we obtain the supercharges and Hamiltonian which still obey the $\cN{=}\,2M$ superalgebra \p{N4SP} and contain only the
surviving pair $(r_n,\phi_n)$ of the originally $2n$ ``angular'' variables.
One may check that the supercharges $Q^a$ and $\bQ_b$ and the Hamiltonian $H$, with the generators $\ell_{ij}$
replaced by ${\hat\ell}_{ij}$ and with the constraints \p{solcon} and \p{gf1} taken into account,
perfectly commute with $r^2_n-\Pi_{nn}$. Thus, the final step of the reduction is to impose the constraint
\be\label{red1}
r^2_n-\Pi_{nn} \approx \mathrm{const} =: g
\ee
and to fix the remaining U(1) gauge symmetry via
\be\label{red2}
\phi_n\approx 0\ .
\ee
The previous two relations are the supersymmetric analogs of \p{redB}.
We conclude that the full set of the reduction constraints reads
\be\label{fincon}
r_i^2 \approx g+\Pi_{ii} \und \phi_i\approx 0 \qquad\mathrm{for}\quad i=1,\ldots,n\ .
\ee

With these constraints taken into account, our supercharges $Q^a$ and $\bQ_b$ and the Hamiltonian $H$ acquire the form
\be\label{finQH}
\begin{aligned}
{\widehat Q}^a \ &=\ \sum_{i=1}^n p_i \psi^a_i\ -\ \im \sum_{i \neq j}^n \frac{\left(\sqrt{g+\Pi_{ii}}\sqrt{g+\Pi_{jj}}-\Pi_{ij}\right) \rho^a_{ji}}{x^i-x^j}\ , \\
{\widehat \bQ}_b \ &=\ \sum_{i=1}^n p_i \bpsi_{i\,b}\ +\ \im \sum_{i \neq j}^n \frac{\brho_{ij\,b}\left( \sqrt{g+\Pi_{ii}}\sqrt{g+\Pi_{jj}}-\Pi_{ji}\right) }{x^i-x^j}\ , \\
\widehat{H} \ &=\ \frac{1}{2}\sum_{i=1}^n p_i^2\ +\ \frac{1}{2}\sum_{i \neq j}^n \frac{\left( \sqrt{g+\Pi_{ii}}\sqrt{g+\Pi_{jj}}-\Pi_{ij}\right)
\left( \sqrt{g+\Pi_{ii}}\sqrt{g+\Pi_{jj}}-\Pi_{ji}\right) }{\left(x^i-x^j\right)^2}\ .
\end{aligned}
\ee
It is matter of quite lengthy and tedious calculations to check that these supercharges and Hamiltonian form an $\cN{=}\,2M$ superalgebra~\p{N4SP}. 
The main complication arises from the expressions $\sqrt{g+\Pi_{ii}}$ present in the supercharges and the Hamiltonian. 
Due to the nilpotent nature of $\Pi_{ij}$, the series expansion eventually terminates, 
but even in the two-particle case with $\Nf$ supersymmetry we encounter a lengthy expression,
\be
\sqrt{g+\Pi_{11}}=\sqrt{g}\,\bigl( 1 +\sfrac{1}{2 g} \Pi_{11}-\sfrac{1}{8 g^2} \Pi_{11}^2+\sfrac{1}{16 g^3} \Pi_{11}^3-
\sfrac{5}{128 g^4} \Pi_{11}^4\bigr)\ .
\ee
For $n$ particles the series will end with a term proportional to $\left(\Pi_{ii}\right)^{\cN (n-1)}$. 
Clearly, these terms will generate higher-degree monomials in the fermions, both for the supercharges and for the Hamiltonian.
We can only speculate that the dread of such complexities impeded an earlier discovery of genuine $\Nf$ Calogero models.

\newpage

\subsection{Simplest example: $\cN{=}\,2$ supersymmetric two-particle Calogero model}
For $\cN{=}\,2$ supersymmetry one has to put $M=1$ in the expressions \p{finQH} for the supercharges and Hamiltonian.
This somewhat reduces their complexity compared to the $\Nf$ case,
but the real simplification occurs for two particles. Indeed, for $n{=}2$ we get
\be\label{simpl1}
\Pi_{22}=-\Pi_{11} \und \Pi_{11}^3  \equiv 0 \qquad \Rightarrow \qquad 
\sqrt{g+\Pi_{11}}\sqrt{g-\Pi_{11}} = \bigl( g -\sfrac{1}{2 g} \Pi_{11}^2\bigr) \qquad\textrm{for}\quad g\neq0\ .
\ee
Moreover, the term $\Pi_{11}^2$ is of the maximal possible power in the $\rho$ and $\brho$ fermions and, therefore, 
disappears from the supercharges. Thus, we are left with
\be\label{N2finQ2}
{\widehat Q}_{(2)}= \sum_{i=1}^2 p_i \psi_i -\im \sum_{i \neq j}^2 \frac{\left(g-\Pi_{ij}\right) \rho_{ji}}{x^i-x^j} \und
{\widehat \bQ}_{(2)}= \sum_{i=1}^2 p_i \bpsi_i + \im \sum_{i \neq j}^2 \frac{\brho_{ij}\left( g-\Pi_{ji}\right) }{x^i-x^j}\ ,
\ee
which have the standard structure -- linear and cubic in the fermions.
The Hamiltonian $\widehat{H}_{(2)}$ reduces to
\be\label{N2Ham2}
\widehat{H}_{(2)}=\frac{1}{2}\sum_{i=1}^2 p_i^2 + \frac{1}{2}\frac{g^2-\Pi_{11}^2- g\left(\Pi_{12}+\Pi_{21}\right)+
 \Pi_{12}\Pi_{21}}{\left(x^1-x^2\right)^2}\ ,
\ee
with the explicit expressions
\be
\Pi_{11}= \rho_{12} \brho_{21}+\brho_{12} \rho_{21}\ , \quad
\Pi_{12}=\left( \psi_1{-}\psi_2\right)\brho_{12}+\left( \bpsi_1{-}\bpsi_2\right)\rho_{12}\ , \quad
\Pi_{21}=\left( \psi_2{-}\psi_1\right)\brho_{21}+\left( \bpsi_2{-}\bpsi_1\right)\rho_{21}\ .
\ee

This $\cN{=}\,2$ supersymmetric two-particle Calogero model has been previously constructed and analyzed in~\cite{Iv1} 
(for details see the review~\cite{Iv3}). This demonstrates that our approach perfectly reproduces the unique known $\cN{=}\,2$ example.

\setcounter{equation}{0}
\section{Conclusion}
We propose a novel $\cN$-extended supersymmetric $su(n)$ spin-Calogero model as a direct supersymmetrization of the bosonic $su(n)$ model~\cite{GS}. 
In the case of $\Nf$ supersymmetry, our model resembles the one constructed in~\cite{Iv1,Iv2}. However, there are two main differences:
\\[-16pt]
\begin{itemize}
\addtolength{\itemsep}{-6pt}
\item the center of mass is free
\item the $su(n)$ generators are not specified in a particular realization.
\end{itemize}
Thanks to these features, we were able to generalize the reduction procedure to the no-spin Calogero model from $\Nf$ supersymmetry 
to any number $\cN{=}\,2M$ of supersymmetries. 
This lead to the discovery of a genuine $\cN{=}\,2M$ supersymmetric rational Calogero model for any number of particles.

Our models belong to same class which was proposed in~\cite{Iv1,Iv2}. Its main features are
\\[-16pt]
\begin{itemize}
\addtolength{\itemsep}{-6pt}
\item a huge number of fermionic coordinates, namely $\cN n^2$ in number rather than the $\cN n$ to be expected 
\item the supercharges and the Hamiltonian contain terms which a fermionic power much larger than three.
\end{itemize}
Clearly, these features merit a more careful and detailed analysis.

The following further developments come to mind:
\\[-16pt]
\begin{itemize}
\addtolength{\itemsep}{-6pt}
\item a superspace description of the constructed models, at least for $\cN{=}\,2$ and $\Nf$ supersymmetry, 
presumably with nonlinear chiral supermultiplets
\item an extension to the Calogero--Sutherland inverse-sine-square model
\item an extension to the Euler--Calogero--Moser system \cite{w} and its reduction to the goldfish system~\cite{gf},
yielding a supersymmetric goldfish model upon reduction, to be compared with recent results from~\cite{Gal1}.
\end{itemize}

\subsection*{\bf Acknowledgements}
We are grateful to Sergey Fedoruk for stimulating discussions. This work was partially supported by the Heisenberg-Landau program. 
The work of S.K.\ was partially supported by Russian Science Foundation grant 14-11-00598, the one of A.S.\ by RFBR grants 18-02-01046 and 18-52-05002 Arm-a. 
This article is based upon work from COST Action MP1405 QSPACE, supported by COST (European Cooperation in Science and Technology).

\end{document}